\documentclass[epj]{svjour}

\usepackage{graphicx}
\usepackage{dcolumn}   
\usepackage{bm}        
\usepackage{amssymb}   
\usepackage{float}
\usepackage{epsfig}

\usepackage{hyperref}
\usepackage{t1enc}
\usepackage[latin2]{inputenc}
\usepackage{color,caption,subcaption}
\usepackage{cite}

\newcommand{\avr}[1]{\left\langle #1\right\rangle}
\definecolor{!R}{rgb}{1,0,0}

\begin{document}

\title{Elliptic flow due to radiation in heavy-ion collisions}

\author{T. S. Bir\'o\inst{1} \and M. Horv\'ath\inst{1} \and Zs. Schram\inst{2}\inst{3}}
\institute{Department of Theoretical Physics, Wigner Research Centre for Physics, Institute for Particle and Nuclear Physics, Budapest, Hungary
\and  Department of Theoretical Physics, University of Debrecen, Debrecen, Hungary 
\and MTA-DE Particle Physics Research Group, Debrecen, Hungary \\
\email{horvath.miklos@wigner.mta.hu}}

\date{Received: date / Accepted: date}

\abstract{
In this paper we demonstrate that radiation patterns could cause flow-like behaviour without any reference to hydrodynamic description. For that purpose we use a statistical ensemble of radiating dipoles, motivated by the investigation of the equivalent photon yield produced by decelerating charges. 
For the elliptic asymmetry factor, $v_2$, we find a reasonable agreement with experimental data.
\PACS{{25.75.Cj}{} \and {24.10.Pa}{} \and {24.10.Nz}{}}
}
\twocolumn
\maketitle

Hydrodynamical simulations are widely used to describe the early-time evolution of proton-nucleus and heavy-ion collisions, see Refs. \cite{hydro_bozek_1,hydro_bozek_2,hydro_bozek_3,hydro_gale_schenke_1,hydro_gale_schenke_2,hydro_shen,hydro_jiang,hydro_romatschke_1,hydro_wiedemann_1,hydro_wiedemann_2,hydro_wiedemann_3} and the references therein. However, the fact that hydrodynamics has a strong predictive power does not imply that it is the only option to explain collective phenomena in such systems. There have been recent efforts to reproduce the flow patterns observed in RHIC and LHC using color scintillating antennas radiating gluons \cite{colorantennas,colorantennas2}. Other authors utilized phenomenological models of color-electric dipoles in order to account for angular correlations in high-energy processes \cite{dipole,dipole2,dipole3}. It is an ongoing debate though, whether a simple effective model, lacking hydrodynamics, could catch the flow-like behaviour or not. Unfortunately, it is rather complicated to explain the collective properties using microscopical models as a starting point. \\
In this letter, our goal is to demonstrate that the radiation originating from a dipole set-up is, in principle, able to match quantitatively the elliptic asymmetry factor $v_2$, measured in heavy-ion experiments. To do so, we discuss the yield of massless particles produced by a decelerating point-like charge. Then we compute the flow coefficient $v_2$ of a dipole composed of two, parallel displaced counter-decelerating charges. Motivated by the formula in Eq. (\ref{v2avr}), we fit various experimental data. Finally, we conclude our analysis discussing several open issues and their relevance for further, more realistic description.

\textit{Radiation produced by decelerating sources.}
According to electrodynamics, an accelerating point charge radiates. One can reinterpret this phenomenon quasi-classically as the emission of photons. It is straightforward to calculate the differential yield of emitted photons when the charge accelerates uniformly on a straight line \cite{unruhgamma}:
\begin{equation}\label{yield}
Y:=\frac{\mathrm{d}N}{k_\perp\mathrm{d}k_\perp\mathrm{d}\eta\mathrm{d}\psi} 
=\frac{2\alpha_\mathrm{EM}}{\pi}\frac{1}{k_\perp^2} \left| 
\, \int\limits_{w_1}^{w_2}\mathrm{d}w\frac{e^{iw\frac{k_\perp}{g}}}{(1+w^2)^\frac{3}{2}} \right|^2.
\end{equation}
Here the azimuth angle $\psi$ is measured in the perpendicular plane, $\eta$ is the rapidity of the emitted photon in the laboratory frame, related to the detecting angle as $\tanh\eta=\cos\theta$, $k_\perp$ is the magnitude of the transverse momentum of the photon. Parameters $w_1$ and $w_2$ are the boosted initial and final velocities of the charge in the frame of the observer ($w_i=\gamma_iv_i$), $g$ is the magnitude of the co-moving acceleration ($c=1$). It is noteworthy that Eq. (\ref{yield}) in the $k_\perp\rightarrow 0$ limit reproduces a bell-shaped rapidity distribution, similar to Landau's hydrodynamical model,and also the plateau known from the Hwa--Bjorken scenario, depending whether the accelerating motion of the charge covers a short or large range in rapidity \cite{accelcharge}. \\
Speculating further, we assume that in the case of light particles produced in a heavy-ion collision a significant part of the yield comes from similar, deceleration induced radiation processes.\\
It is also worthwhile to mention that a gauge field theory which describes the radiation phenomena on the microscopic level, can be reformulated in the framework of hydrodynamics, as it was endeavoured in Refs. \cite{jackiw,jackiw2,jackiw3}.

\begin{figure}
\centering
  \includegraphics[width=0.9\linewidth]{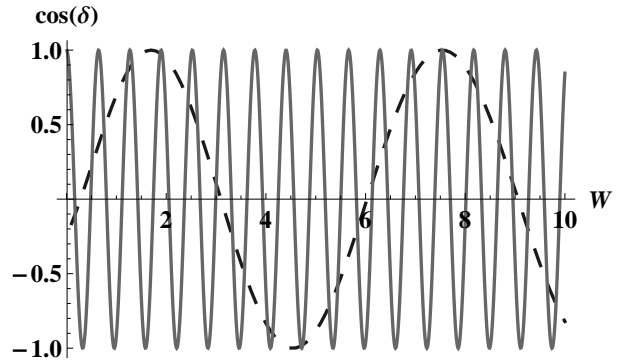}
  \caption{Phase-shift factor $\cos\delta$ as a function of the final velocity $w_2$ in cases $\Delta=1.0$ (dashed line) and $\Delta=10.0$ (solid line).}
\label{fig:phaseshift}
\end{figure}

\textit{Elliptic flow coefficient.} We now turn to the analysis of the simplest structure. An elliptic asymmetry could stem from two decelerating point-like sources going into opposite directions on parallel paths. We calculate the emitted photon-equivalent radiation for the distance $d$ between the two sources. The yield is calculated coherently summing the contributing amplitudes:
\begin{eqnarray}\label{dipoleyield}
Y &\propto & |\mathcal{A}_1+\mathcal{A}_2|^2 =\nonumber \\
&=& |A_1e^{i\frac{d}{2}k_\perp \cos(\varphi-\psi)}+A_2e^{-i\frac{d}{2}k_\perp \cos(\varphi-\psi)}|^2 
\nonumber \\
&=& |A_1|^2+|A_2|^2+2\mathrm{Re}(A_1A_2^* e^{ik_\perp d \cos(\varphi-\psi)}).
\end{eqnarray}
Here $\varphi$ is the angle between the dipole axis and the detector, and $\psi$ characterizes the event-plane. The elliptic flow coefficient we deal with here is defined by the azimuthal averaging of $\cos(2(\varphi-\psi))$ with respect to the normalized yield:
\begin{equation}\label{dipolev2}
v_2= -\frac{2\mathrm{Re}(A_1A_2^*)J_2(k_\perp d)}{|A_1|^2+|A_2|^2+2\mathrm{Re}(A_1A_2^*)J_0(k_\perp d)}.
\end{equation}
In the above formula $J_n$ is the Bessel function of the first kind. It is part of the Jacobi--Anger expansion, $e^{iz\cos(t)} = \sum_{n=-\infty}^{\infty} i^n J_n(z) \, e^{int}$, applied to $e^{i\frac{d}{2}k_\perp \cos(\varphi-\psi)}$. Integrating with the factor $\cos(2(\varphi-\psi))$ and dividing by the zeroth order coefficient delivers Eq. (\ref{dipolev2}). Parametrizing the complex amplitudes as $A_1=Ae^{i\delta_0}$ and $A_2=\gamma Ae^{i(\delta_0+\delta)}$ with real $A$, $\gamma$ and $\delta_0$, where $\delta$ denotes the phase-shift and $\gamma$ is the ratio of $|A_2|/|A_1|$, we obtain the simplified expression
\begin{equation}
 v_2 = \frac{-J_2(k_\perp d)\cos\delta}{\frac{1+\gamma^2}{2\gamma}+J_0(k_\perp d)\cos\delta }.
	\label{v2final}	
\end{equation}
Further we assume that $A_1$ and $A_2$ are azimuthally symmetric, i.e. independent of the difference $\varphi-\psi$.\\
In our picture $v_2$ depends on the phase-shift $\delta$, the dipole size $d$ and the strength asymmetry parameter $\gamma$. We assume that event-by-event $d$ and $\gamma$ might be well-determined, while $\delta$ fluctuates. The decelerating sources are strongly affected by the medium surrounding them, therefore, they radiate differently, depending on how long the interaction holds up. We consider one source decelerating from a velocity near $c$ to $0$ and the other one from $c$ to slightly above 0, in the opposite direction. The relevant amplitudes then read:
\begin{eqnarray}
\mathcal{A}_1 &=& 
\int\limits_{-\infty}^{w_1=0}\mathrm{d}w\frac{e^{iw\Delta}}{(1+w^2)^\frac{3}{2}} =\nonumber \\
&=&\frac{\Delta}{2}\left[ 2K_1(\Delta) +i\pi \left(K_1(\Delta)-L_{-1}(\Delta)\right) \right], 
	\nonumber \\
\mathcal{A}_2 &=& 
\int\limits_{\infty}^{w_2}\mathrm{d}w\frac{e^{iw\Delta}}{(1+w^2)^\frac{3}{2}}
= -\mathcal{A}_1^*+ \int\limits_0^{w_2}\mathrm{d}w\frac{e^{iw\Delta}}{(1+w^2)^\frac{3}{2}},
\end{eqnarray}
as it follows from Eq. (\ref{yield}) with $\Delta=\frac{k_\perp}{g}$, $K_n$ and $L_\nu$ being the modified Bessel functions of second kind and the modified Struve function, respectively. The phase-shift factor, $\cos\delta=\frac{\mathrm{Re}(\mathcal{A}_1\mathcal{A}_2^*)}{|\mathcal{A}_1||\mathcal{A}_2|}$, can be evaluated numerically, the result is plotted in Fig. \ref{fig:phaseshift}. It appears as a natural idea to average with respect to the phase difference variable, $\delta$, whenever $\cos\delta$ oscillates fast as a function of $w_2$ (cf. Fig. \ref{fig:phaseshift}).

\begin{figure}
	\includegraphics[width=1.0\linewidth]{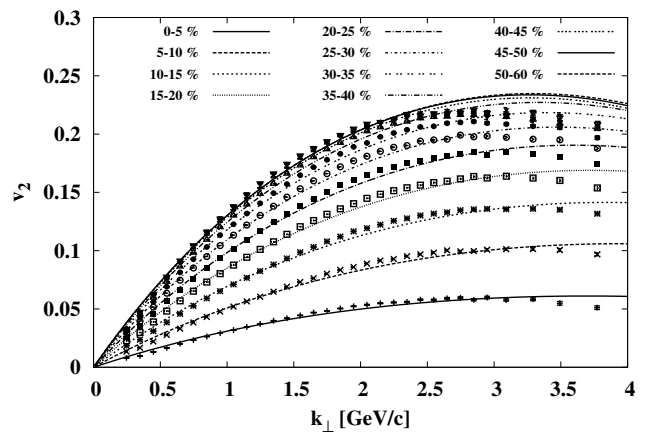}
\caption{Elliptic flow of charged hadrons measured by PHENIX in Au-Au collisions at $\sqrt{s_{NN}}=200$ GeV for several centrality classes \cite{idParts_v2_PHENIX2}}
\label{fig:chhadrons}
\end{figure}

\begin{figure}
\centering
\includegraphics[width=1.0\linewidth]{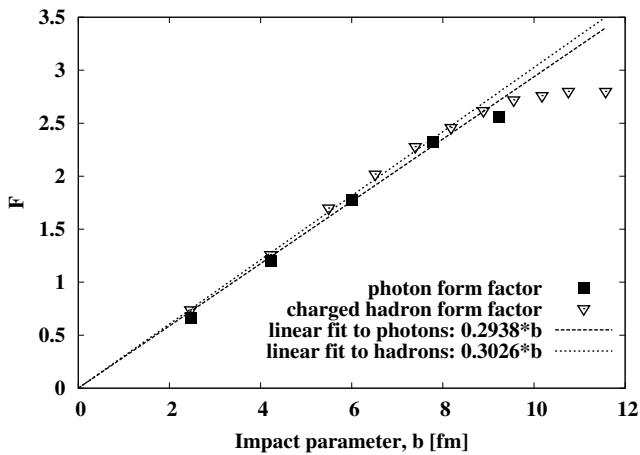}
\caption{The form factor $F$ -- see Eq. (\ref{v2fit}) -- versus the impact parameter $b$ of the collision for inclusive photons and charged hadrons. The centrality--impact parameter relationship is taken from \cite{centr_vs_multip}.}
\label{fig:formfactor}
\end{figure}

\textit{Fits to experimental data.} The uniform averaging respect to the phase-shift angle can be carried out analytically, resulting
\begin{equation}\label{v2avr}
\avr{v_2} = \frac{J_2(k_\perp d)}{J_0(k_\perp d)}\left(\frac{1}{\sqrt{1-\frac{4\gamma^2}{(1+\gamma^2)^2}J_0^2(k_\perp d)}}-1\right).
\end{equation}
Hereinafter we assume that the leading order contribution to the elliptic asymmetry comes from a dipole-like structure discussed previously. We shall test our hypothesis on elliptic flow measurements in heavy-ion collisions of RHIC and LHC, where the fairly large number of dipoles ensures the validity of our working hypothesis, namely the uniform phase-shift of the sources. We introduce an additional fit parameter, called $F$. This geometrical form factor is independent of the transverse momentum $k_{\perp}$. Finally the formula we use to fit the experimental data scales the expression (\ref{v2avr}) as:
\begin{equation}\label{v2fit}
\avr{v_2}_\mathrm{fit} \: = \: F \cdot \avr{v_2}\!\left(k_\perp d,\gamma\right).
\end{equation}
Comparison of measured and calculated $v_2$ as a function of $k_\perp$ are depicted on Figs. \ref{fig:photons}, \ref{fig:pions} and \ref{fig:chhadrons}. For data resolved by centrality the fitted parameter, $d$, remains the same within 11\% for photons and 19\% for charged hadrons, cf. Table \ref{tab:tab1}. Including more peripheral collisions, $F$ saturates somewhat below 3 (see Fig.\ref{fig:formfactor}). Since $F=1$ would mean that only the dipole term contributes to $v_2$, cf. Eq. (\ref{v2avr}), that suggests the need for some further sources of elliptic flow, for example multipole contributions. In all cases $\gamma$ turns out to be very close to one, showing that symmetric dipole sources may dominate the experimental findings. The $v_2$ values for charged pions on Fig. \ref{fig:pions} and hadrons on Fig. \ref{fig:chhadrons} fit as well as for the emitted photons on Fig. \ref{fig:photons}. \\
The typical value of the effective dipole size $d$ is about $0.06$ fm from the hadronic fits and about $0.1$ fm from photon data. This is rather small compared to the size of heavy-ion fireballs, it may hint to subhadronic sources of this part of radiation. We mention here that other authors pointed out the quark-level origin of the flow independently of our present analysis \cite{lacey}. \\
At this point, the physical interpretation of the form factor $F$ is due. Being momentum independent, a simple geometric cartoon of a heavy-ion collision can be suggested. Averaging the yield after Fourier-expansion of Eq. (\ref{dipoleyield}) with respect to a profile parametrized by polar coordinates we obtain 
\begin{equation}
Y=Y_0\avr{1}+Y_2\avr{\frac{1}{\cos 2\varphi}}\cos 2\varphi= Y_0+FY_2\cos 2\varphi,
\end{equation}
with $\varphi$ being the polar angle measured from the reaction plane. In case of an ellipse, the factor in front of $Y_2$ is the anisotropy $F=\frac{A^2-B^2}{A^2+B^2}$, with the half-axes $A$ (in the reaction plane) and $B$ (perpendicular to the reaction plane). There are several ways to attach an ellipse to the geometry of the collision. Let us consider two nuclei as circular disks (squeezed due to Lorentz contraction) with radius $R$, displaced by impact parameter $b$ between the centres. In case of approximating the intersection by an ellipse ($A=R-\frac{b}{2}$, $B^2=R^2-\frac{b^2}{4}$) and assuming subhadronic dipoles ordered perpendicular to the reaction plane we obtain $F=\frac{b}{2R}$. A much larger ellipse including also the spectator area ($A=R+\frac{b}{2}$, $B^2=R^2-\frac{b^2}{4}$) with dipoles ordered parallel to the reaction plane delivers the same anisotropy $F=\frac{b}{2R}$. In fact experimentally $F(b)$ turns out to be linear in a wide range of impact parameter values, see Fig. \ref{fig:formfactor}. For fitting the anisotropy formula mentioned above we get $R\approx1.67$fm. This is an effective size of the source of the dipole-like radiation. It is smaller than the typical size of a Pb-nucleus by a factor of four. This finding warns against a collective source extending in the whole media, but does not exclude hydrodynamic evolution.
{\centering
\begin{figure}
	\includegraphics[width=1.0\linewidth]{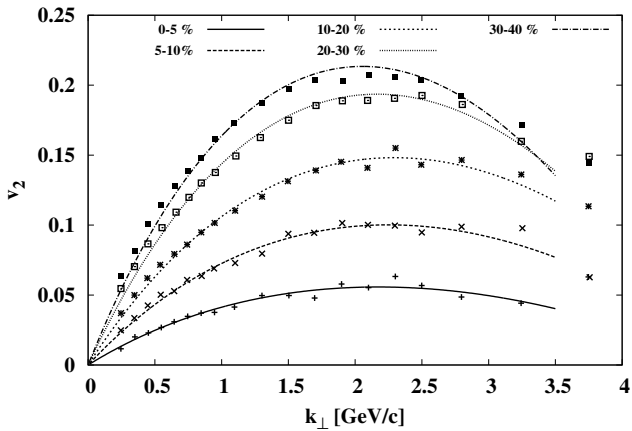}
\caption{Inclusive photon elliptic flow measured by \mbox{ALICE} group of LHC in Pb-Pb collisions at $\sqrt{s_{NN}}=2.76$ TeV for several centrality classes \cite{inclPhoton_v2_ALICE}}
\label{fig:photons}
\end{figure}
\begin{figure}
	\includegraphics[width=1.0\linewidth]{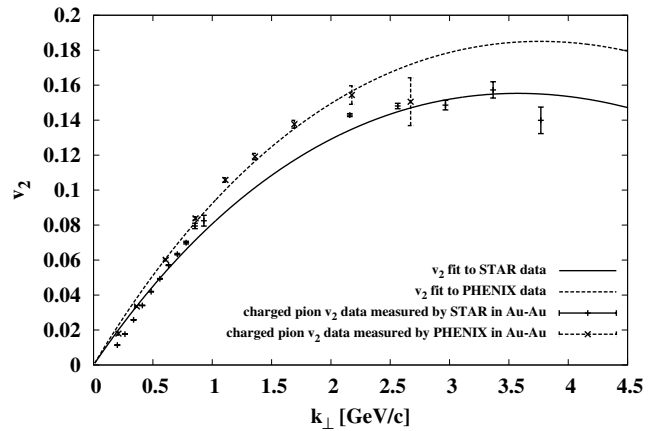}
\caption{Elliptic flow of charged pions measured by STAR of RHIC in Au-Au collisions at $\sqrt{s_{NN}}=62.4$ GeV (solid line) \cite{ideParts_v2_STAR}, and by PHENIX of RHIC in Au-Au collisions at $\sqrt{s_{NN}}=200$ GeV (dashed line) \cite{idParts_v2_PHENIX}}
\label{fig:pions}
\end{figure}
}

\textit{Discussion and conclusions. }
We briefly list recent literature studies about how various stages of the heavy-ion collision could contribute to the azimuthal asymmetry of the flow in order to support the phenomenological picture we sketched above. We focus the role of dipole-like structures revealing at the early-time stage of the heavy-ion collisions.\\
\textit{i.) Strong electromagnetic (EM) fields.} In non-central collisions, the magnitude of the magnetic field due to the geometrical asymmetry of the system could reach $\sim 5 m^2_\pi$ for a short time of $0.2$fm/c \cite{HIJING_EM,HSD_EM}. The pure EM-effect (caused by the coupling of charged quasi-particles and the EM-field) is, however, not significant at the level of global observables, as it is suggested by hadron string dynamics simulations \cite{HSD_EM}, or contributes to higher order asymmetries only (quadrupole electric moment) \cite{HIJING_EM}. Note that these simulations are based on transport models using quasi-particles and improved, but essentially perturbative cross sections.\\
In Ref. \cite{mclerran_skokov} the authors use an order-of-magnitude estimation, leading order in perturbation theory, for the gluon-photon coupling in order to argue that the direct photon flow maybe affected at RHIC, but unlikely at LHC.\\
\textit{ii.) QCD in magnetic field.} Lattice Monte-Carlo simulations suggest that QCD at high temperatures is paramagnetic Ref. \cite{latticeQCD}. Therefore a 'squeezing' of the plasma could occur, elongating it in the direction of external magnetic field, which, in case of non-central collisions, points perpendicular to the reaction plane. Charge separation of quarks in the direction of the external magnetic field due to the fluctuation of the topological charge (known as the chiral magnetic effect, CME) can also contribute to the asymmetry of the plasma, as it is indicated by lattice results \cite{latticeQCD2}. These effects are not yet incorporated in quasi-particle simulations like in Refs. \cite{HIJING_EM,HSD_EM}.\\
\textit{iii.) Radiation of non-Abelian plasma.} The classical limit of non-Abelian fields generated by ultra-relativistic sources is analysed in Ref. \cite{SU2dipole}. It is shown that dipole-like structures will emerge with the same geometric properties like their EM-versions. These could be important when initial state of the matter like the color glass condensate (CGC) is melting down and converting to QGP, while a considerable amount of quark-antiquark pairs are produced. This happens probably when dipoles are smaller than 1 fm, accompanied by fast oscillation of the sources.\\
It is pointed out in Ref. \cite{Goloviznin} that the bremsstrahlung of quarks on the surface of QGP, pulled back by the confining force could produce photon radiation in comparable amount to those produced in the plasma phase.
An important aspect of the issue of the azimuthal asymmetries is in what extent are those evolved on the microscopic level of the dynamics or caused by collective behaviour. There is an ongoing debate in the literature about the contributions of initial and final state asymmetries to the elliptic flow \cite{Goloviznin,lappi,schenke,rybicki}. Emphasizing only a few examples, it is observed that in proton-nucleon collisions the CGC-correlations could be directly visible in the measurable particle spectrum. Using classical Yang-Mills simulations for p-Pb collisions, in the first half fm/c after CGC was initiated, significant build-up of contributions to $v_2$ and $v_3$ was observed \cite{lappi}. These momentum space anisotropies are not correlated with the final state global asymmetries described as collective flow behaviour. In Pb-Pb collisions, the early-time contributions are relatively small, supporting the role of collective effects. In this case the sources are uncorrelated, localized color field domains, resulting the gluon spectrum to be isotropic \cite{schenke}. EM effects also could play a role in the final state. The directed flow of charged pions could be a result of a spectator induced splitting, as it is demonstrated in Refs. \cite{rybicki,rybicki2}. \\
It seems that dipole-like structures coupled to the initial geometric asymmetry of heavy-ion collisions are quite natural in a wide scale of models concerning the early time evolution of the hot nuclear matter. We suggest that these domains could be the sources of intense gluon and/or photon radiation having similar geometric properties to its EM counterparts. The orientation of these dipoles may be ordered by EM effects like the mentioned squeezing of the QCD plasma and CME, triggered by the early-time intense fields present in non-central events. Therefore the cumulative effect of small but not necessarily coherent radiators may affect the macroscopic observables, contributing  significantly to the azimuthally asymmetric component of the flow. \\
In conclusion, we emphasize the necessity of exploring how microscopic causes can lead to macroscopic anisotropies. Especially in large systems, where collective (flow-like) effects may take place, it is important to distinguish those from the individual subhadronic causes.
\vspace{\baselineskip}\\
\textit{Acknowledgement.} Discussions on gluon bremsstrahlung with M. Gyulassy and on the radiation patterns with Zs. Szendi are gratefully acknowledged. We thank M. Vargyas and G. G. Barnaf\"oldi for critically reading the manuscript. This work has been supported by the Hungarian National Science Fund, OTKA K 104260, by the Chinese-Hungarian bilateral cooperation NIH TET 12CN-1-2012-0016 and by the Helmholtz International Center for FAIR within the LOEWE program launched by the State of Hesse.
\begin{table}[h]\centering
\begin{tabular}{c|c|c|c|c}
\hline
 Cent. [\%] & $F$ & $d$ [fm] & $\gamma$ & $\chi^2$ \\
\hline\hline
\multicolumn{5}{c}{ALICE photon $v_2$ fit parameters}\\ \hline
0$-$5&	 0.66&	0.10& 1.00&	0.36\\
5$-$10&	 1.20&	0.10	& 1.00&	0.36\\
10$-$20&	 1.78&	0.09	& 1.00&	0.58\\
20$-$30&	 2.32&	0.10	& 1.00&	0.82\\
30$-$40&	 2.56&	0.10	& 1.00&	1.60\\
\hline
\multicolumn{5}{c}{PHENIX charged hadron $v_2$ fit parameters} \\ \hline
0$-$5&	 0.74&	0.06& 1.00&	4.46\\
5$-$10&	 1.26&	0.05	& 1.00&	4.31\\
10$-$15&	 1.70&	0.06	& 1.00&	5.25\\
15$-$20&	 2.02&	0.06	& 1.00&	6.09\\
20$-$25&	 2.28&	0.06	& 1.00&	6.88\\
25$-$30&	 2.46&	0.06	& 1.00&	7.24\\
30$-$35&	 2.62&	0.06	& 1.00&	7.55\\
35$-$40&	 2.72&	0.06	& 1.00&	7.91\\
40$-$45&	 2.76&	0.07	& 1.00&	8.19\\
45$-$50&	 2.80&	0.07	& 1.00&	7.53\\
50$-$60&	 2.80&	0.07	& 1.00&	7.57\\
\hline
\multicolumn{5}{c}{STAR charged pion $v_2$ fit parameters} \\ \hline
& 1.86&	0.06&	1.00&	4.48\\
\hline
\multicolumn{5}{c}{PHENIX charged pion $v_2$ fit parameters} \\ \hline
& 2.22&	0.06&	1.00&	0.15
\end{tabular}
\caption{List of fit parameters. $\chi^2$ values contain the published measurement errors and also the error of fitting.}
\label{tab:tab1}
\end{table}


\begin{thebibliography}{100}
\bibitem{hydro_bozek_1} P. Bo\.{z}ek, W. Broniowski, arXiv:1403.6042v1 
\bibitem{hydro_bozek_2} P. Bo\.{z}ek, Physics Letters B 699, 283-286 (2011) 
\bibitem{hydro_bozek_3} P. Bo\.{z}ek, I. Wyskiel-Piekarska, Phys. Rev. C 85, 064915 (2012) 
\bibitem{hydro_gale_schenke_1} C. Gale, S. Jeon, B. Schenke, P. Tribedy, R. Venugopalan, Phys. Rev. Lett. 110, 012302 (2013) 
\bibitem{hydro_gale_schenke_2} B. Schenke, S. Jeon, C. Gale, Physics Letters B 702, 59-63 (2011) 
\bibitem{hydro_shen} Chun Shen, U. Heinz, P. Huovinen, H. Song, Phys. Rev. C 84, 044903 (2011) 
\bibitem{hydro_jiang} Z. J. Jiang, Q. G. Li, H. L. Zhang, Phys. Rev. C 87, 044902 (2013) 
\bibitem{hydro_romatschke_1} P. Romatschke and U. Romatschke, Phys. Rev. Lett. 99, 172301 (2007) 
\bibitem{hydro_wiedemann_1} R. Baier, P. Romatschke, U. A. Wiedemann, Phys. Rev. C 73, 064903 (2006), arXiv:hep-ph/0602249 
\bibitem{hydro_wiedemann_2} S. Floerchinger, U. A. Wiedemann, A. Beraudo, L. D. Zanna, G. Inghirami, V. Rolando, Phys. Lett. B735, 305-310 (2014), arXiv:1312.5482 
\bibitem{hydro_wiedemann_3} S. Floerchinger, U. A. Wiedemann, A. Beraudo, L. D. Zanna, G. Inghirami, V. Rolando, Nucl. Phys. A931 965-969 (2014), arXiv:1408.6384 

\bibitem{colorantennas} M. Gyulassy, P. Levai, I. Vitev, T. S. Bir\'o, Phys.Rev. D 90 054025 (2014), arXiv:1405.7825 
\bibitem{colorantennas2} M. Gyulassy, P. Levai, I. Vitev, T. S. Bir\'o, Nucl. Phys. A 931 943-948 (2014), arXiv:1407.7306  
\bibitem{dipole} A. Kovner, M. Lublinsky, Phys. Rev. D 84, 094011 (2011) 
\bibitem{dipole2} A. Dumitru, A. V. Giannini, Nuclear Physics A 933 212 (2015), arXiv:1406.5781 
\bibitem{dipole3} A. Dumitru, L. McLerran, V. Skokov, arXiv:1410.4844 
\bibitem{unruhgamma} T. S. Bir\'o, M. Gyulassy, Z. Schram, Phys. Lett. B 708, 276-279 (2012), arXiv:1111.4817 
\bibitem{accelcharge} T. S. Bir\'o, Z. Szendi, Z. Schram, Eur. Phys. J. A 50, 60 (2014), arXiv:1401.1987 

\bibitem{jackiw} R. Jackiw, S.-Y. Pi, A. P. Polychronakos, Annals Phys. 301, 157-173 (2002) arXiv:hep-th/0206014v2 
\bibitem{jackiw2} R. Jackiw, arXiv:hep-th/0305027v2 (2003) 
\bibitem{jackiw3} R. Jackiw, arXiv:hep-th/0410284v1 (2004) 

\bibitem{lacey} R: Lacey, A. Taranenko, PoSCFRNC2006:021 (2006), arXiv: nucl-ex/0610029v3 

\bibitem{inclPhoton_v2_ALICE} D. Lohner \textit{et al.} (ALICE Collaboration), J. Phys. Conf. Ser. 446, 012028 (2013) 
\bibitem{ideParts_v2_STAR} B. I. Abelev \textit{et al.} (STAR Collaboration), Phys. Rev. C 75, 054906 (2007) 
\bibitem{idParts_v2_PHENIX} A. Adare \textit{et al.} (PHENIX Collaboration), Phys. Rev. Lett. 91, 182301 (2003) 
\bibitem{idParts_v2_PHENIX2} A. Adare \textit{et al.} (PHENIX Collaboration), Phys. Rev. Lett. 105, 062301 (2010), arXiv:1003.5586v2 
\bibitem{centr_vs_multip} B. Abelev \textit{et al.} (ALICE Collaboration), Phys. Rev. C 88, 044909 (2013)

\bibitem{HIJING_EM} Wei-Tian Deng, Xu-Guang Huang, Phys. Rev. C 85, 044907 (2012) 
\bibitem{HSD_EM} V. Voronyuk, V. D. Toneev, W. Cassing, E. L. Bratkovskaya, V. P. Konchakovski, S. A. Voloshin, Phys. Rev. C 83, 054911 (2011) 
\bibitem{mclerran_skokov} L. McLerran, V. Skokov, report No. BNL-100762-2013-JA, arXiv: 1305.0774v1 
\bibitem{latticeQCD} G. S. Bali, F. Bruckmann, G. Endr\H{o}di, and A. Sch\"afer, Phys. Rev. Lett. 112, 042301 (2014) 
\bibitem{latticeQCD2} G. S. Bali, F. Bruckmann, G. Endr\H{o}di, Z. Fodor, S. D. Katz and A. Sch\"afer, JHEP 1404 129 (2014), arXiv: 1401.4141 


\bibitem{SU2dipole} W. Cassing, V. V. Goloviznin, S. V. Molodtsov, A. M. Snigirev, V. D. Toneev, V. Voronyuk, and G. M. Zinovjev, Phys. Rev. C 88, 064909 (2013) 
\bibitem{Goloviznin} V. V. Goloviznin, A. M. Snigirev, G. M. Zinovjev, Pis'ma v ZhETF, vol. 98, iss. 2, pp. 69-71 (2013), arXiv:1209.2380 

\bibitem{lappi} T. Lappi, arXiv: 1501.05505 (2015) 
\bibitem{schenke} B. Schenke, S. Schlichting, R. Venugopalan, arXiv: 1502.01331 (2015) 
\bibitem{rybicki} A. Rybicki and A. Szczurek, Phys. Rev. C 87, 054909 (2013), arXiv:1303.7354v1 
\bibitem{rybicki2} A. Rybicki, A. Szczurek, M. Klusek-Gawenda, M. Kielbowicz, Conference: C14-08-25.8, arXiv:1502.03689 


\end{thebibliography}
\end{document}